\documentclass[superscriptaddress,twocolumn,aps,prb,floatsfix]{revtex4}
\usepackage{graphicx}
\usepackage{amssymb,amsmath}
\usepackage{array}
\usepackage{dcolumn}

\begin{document}
 
\title{An ab-initio evaluation of the local effective interactions in the
superconducting compound $\rm Na_{0.35} Co O_2-1.3H_2O$}
 
\author{Sylvain Landron}
\author{Marie-Bernadette Lepetit}
 
\affiliation{CRISMAT, ENSICAEN-CNRS UMR~6508, 6~bd. Mar\'echal Juin,
14050 Caen, FRANCE}

\date{\today}

\begin{abstract}
We used ab-initio quantum chemical methods, treating explicitly the
strong correlation effects within the cobalt $3d$ shell, as well as
the screening effects on the effective integrals, for accurately
determining on-site and nearest-neighbor (NN) interactions in the $\rm
Na_{0.35} Co O_2-1.3H_2O$ superconducting compound. The effective
ligand field splitting within the $t_{2g}$ orbitals was found to be
$\delta \sim 300\,\rm meV$, the $a_{1g}$ orbital being destabilized
compared to the $e_g^\prime$ ones. The effective Hund's exchange and
Coulomb repulsion were evaluated to $J_H\sim 280\, \rm meV$ and $U\sim
4.1$--$4.8\, \rm eV$ for the $a_{1g}$ orbitals. The NN hopping
parameters were determined within the three $t_{2g}$ orbitals and
found to be of the same order of magnitude as the $t_{2g}$ ligand
field splitting. This result supports the hypothesis that a three band
model would be better suited than a one-band model for this
system. Finally we evaluated the NN effective exchange integral to be
antiferromagnetic and $J=-66\,\rm meV$.
\end{abstract}
\maketitle                                                                            

\section{Introduction}
The layered cobalt oxides have been attracting a lot of attention for
the last few years. This interest is driven by the remarkable
properties of the $\rm Na_x Co O_2$ compounds and more specially their
hydrated counterpart. Indeed, superconductivity was discovered in the
$\rm Na_{0.35} Co O_2-1.3H_2O$~\cite{supra}, for the first time in
layered oxides, beside its discovery in the cuprates.

The host material is composed of $\rm Co O_2$ layers, where the cobalt
atoms are located in distorted edge-sharing octahedra forming a
two-dimensional triangular lattice. The sodium atoms are located in a
plane in between the $\rm Co O_2$ layers. In the hydrated
superconducting material, the water molecules are intercalated in
between the sodium and the cobalt oxide layers. The water intercalation is
fragile and the system looses its water out of a hydrated atmosphere.
It seems however, that at low temperature, the water and sodium cations
order in a two-dimensional super-cell and adopt a frozen local geometry
similar to clusters of $\rm Na^+$ ions embedded in ice~\cite{ordre}.
Despite this possible ordering, it is believed that the effect of the
water molecules is only steric~\cite{supra}. Indeed, the water
inclusion induces a large separation of the $\rm Co O_2$ layers,
responsible for essentially uncoupled cobalt layers and a
two-dimensional physics~\cite{eau}. This two-dimensional character is
assumed to be necessary for superconductivity to occur.

A simple formal charge analysis shows that the cobalt ions are $3.65+$
for $x=0.35$, that is  about one third of $\rm Co^{3+}$ ions
and two third of $\rm Co^{4+}$ ions. Wet-chemical redox analyses
revealed however a cobalt oxidation number somewhat
lower~\cite{oxyd}~: 3.46+, that is closer to half $\rm Co^{3+}$, half
$\rm Co^{4+}$ ions.
The distortion of the $\rm Co O_6$ octahedra observed in the
superconducting material corresponds to a compression along the
$x+y+z$ axis of the $\rm Co$ coordination octahedron. This trigonal
distortion induces a lowering of the $O_h$ local point group to a
$d_{3}$ subgroup, thus splitting the $t_{2g}$ orbitals in a $a_{1g}$
and two $e_g^\prime$ ones (see figure~\ref{f:split} in the results
section). Authors however disagree on the relative energies of the
orbitals and the amplitude of the splitting. While some
authors~\cite{Mae} support the idea that the $a_{1g}$ orbital is
lowered compared to the two $e_g^\prime$, other authors come to the
opposite conclusion~\cite{bask}. Density functional Theory (DFT)
calculations~\cite{DFT1} agree on the fact that the $a_{1g}$ band is
less filled than the ones originating from the $e_g^\prime$ orbitals.
However, the relative energies of the different atomic configurations
are not directly accessible to DFT calculations.  Indeed, electronic
correlation is assumed to be very strong in this system and can be
expected to strongly influence the local atomic excitation energies
between the different configurations associated with the hole
localization either on the $a_{1g}$ or on of the two $e_g^\prime$
cobalt atomic orbitals.  The question of the relative energy of these
different configurations is however crucial in order to determine the
pertinent degrees of freedom to be taken into account in a simple
model, able to describe the low energy properties of the
system. Indeed, assuming that the $a_{1g}$ orbital is much higher in
energy that the $e_g^\prime$ ones, one should naturally conclude that
the pertinent model for the description of the superconductivity is a
one band $t-J$ type of model. Assuming that the $a_{1g}$ orbital is
now much lower in energy that the $e_g^\prime$ ones, one comes to a
two-band model, while if the three orbitals are only weakly split, the
pertinent model should consider all of them at the time. No consensus
is reached nowadays in the literature and there is a large controversy
on the pertinent model to consider.

The aim of the present paper is to determine the local orbital
energies and effective coupling parameters between the cobalt $3d$
orbitals. For this purpose we used embedded clusters calculations and
quantum chemical ab-initio methods treating exactly the correlation
effects within the $3d$ shell as well as the screening effects that
renormalize the interactions. Such methods allow the direct
computation of the local parameters such as the atomic effective
ligand field splitting, the one-site Hubbard $U$ coulombic repulsion
as well as the Hund's exchange. In addition, nearest neighbor exchange
and transfer interactions can be directly and accurately computed.

The next section will shortly describe the method, section three will
relate the results and finally the last section will be devoted to
discussion and conclusion.

\section{Method and computational details}
The method used in this work (CAS+DDCI~\cite{DDCI}) is a configurations
interaction method, that is an exact diagonalisation method within a
selected set of Slater determinants, on embedded crystal fragments.
This method has been specifically designed to accurately treat
strongly correlated systems, for which there is no single-determinant
description. The main point is to treat exactly all correlation
effects and exchange effects within a selected set of orbitals (here
the $3d$ shell of the cobalt atoms) as well as the excitations
responsible for the screening effects on the exchange, repulsion,
hopping, etc.  integrals.

The CAS+DDCI method has proved very efficient to compute, within
experimental accuracy, the local interactions (orbital energies, atomic
excitations, exchange and transfer integrals, coulomb repulsion etc.)
of a large family of strongly correlated systems such as high $T_c$
copper oxides~\cite{DDCIhtc}, vanadium oxides~\cite{vana}, nickel and
cuprate fluorides~\cite{DDCI_Ni}, spin chains and
ladders~\cite{incom}, etc.

\begin{figure}[h] 
\resizebox{!}{4cm}{\includegraphics{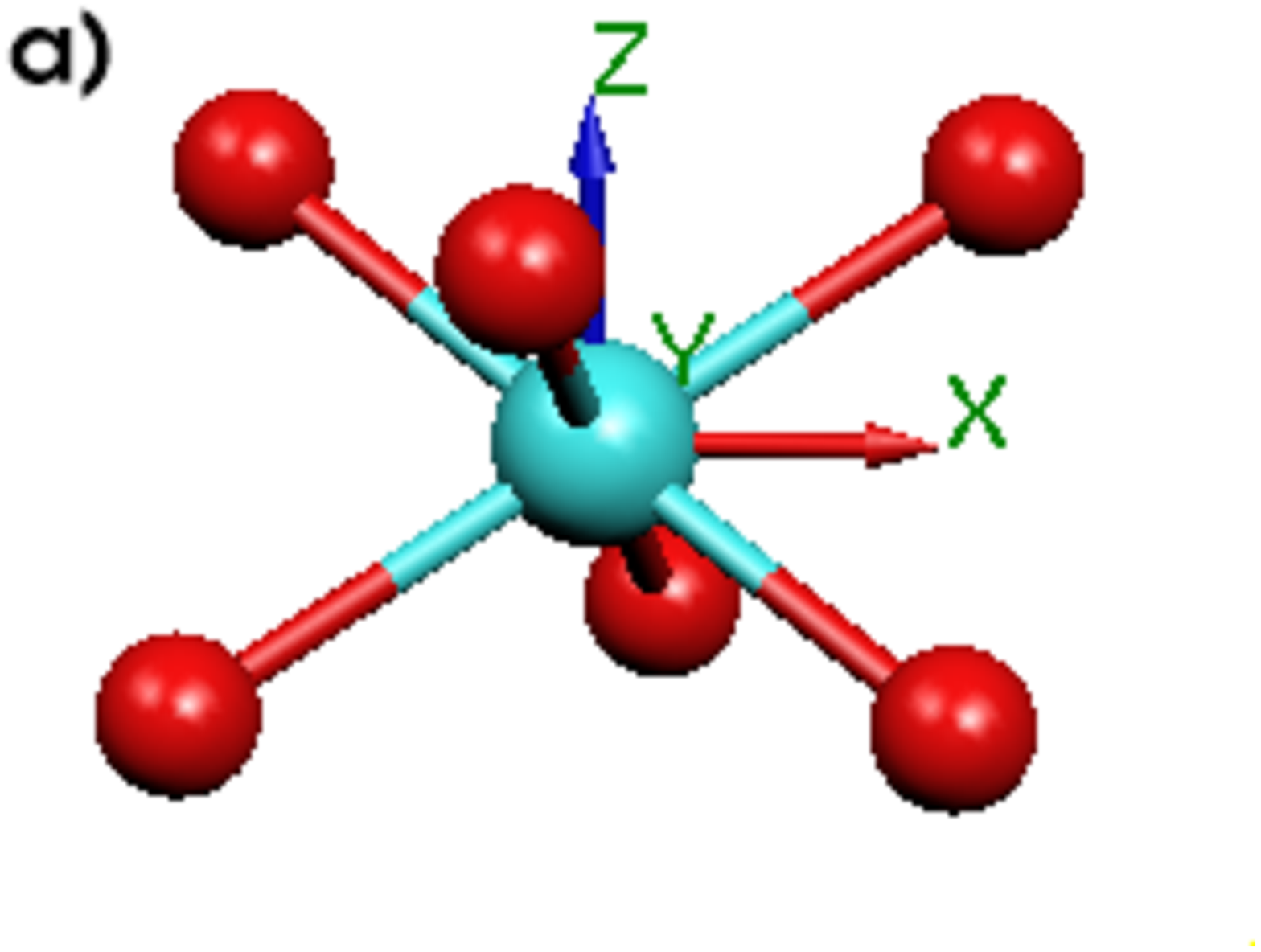}} \\
\resizebox{!}{4cm}{\includegraphics{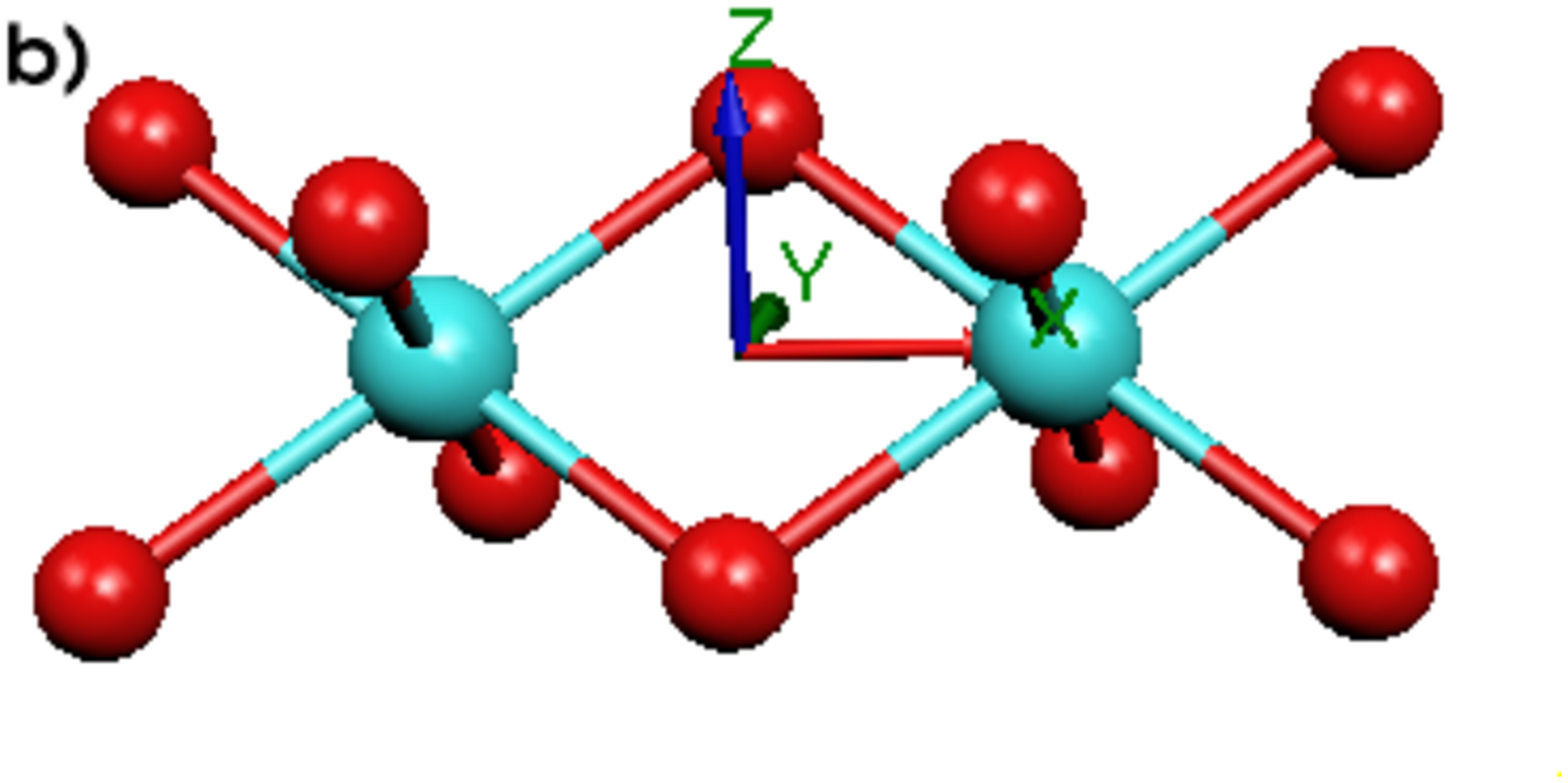}}
\caption{a) $\rm CoO_6$ and b) $\rm Co_2 O_{10}$ clusters used in the
present  calculations.}
\label{f:clus}
\end{figure}

The clusters used in this work involve either one cobalt ($\rm Co
O_6$) or two cobalt atoms ($\rm Co_2 O_{10}$) and their oxygen first
coordination shell (see figure~\ref{f:clus}). These fragments are
embedded in a bath designed so that to reproduce on them the main
effects of the rest of the crystal~; that is the Madelung potential
and the exclusion effects of the electrons of the other atoms of the
crystal on the clusters electrons.

The electrostatic potential is reproduced by a set of point charges
located at the atomic positions. The charges are renormalized next to
the bath borders in order to obtain an exponential convergence of the
Madelung potential. The convergence accuracy was set in the present
work to the mili-electron-Volt. The method used for this purpose is a
generalization~\cite{alain} of the Evjen's method~\cite{Evjen}. The
nominal atomic charges used in this work are the formal charges, that
is $+3.65$ for the cobalt atoms, $-2$ for the oxygen atoms and $+1$
for the sodium atoms. The sodium atoms being located at two
crystallographic sites, with fractional occupations, we renormalized
the associated charges with the crystallographic occupation, thus
using a mean field averaging of the Madelung potential.  At this point
we would like to shortly discuss the question of the $\rm Co$ valency
and insertion of $\rm [H_3O]^+$ ions. Indeed, it is clear that both
the $\rm Co^{3+}$ and $\rm Co^{4+}$ atomic configurations of the Co
atom are present in the system, however the actual average charge
supported by the cobalt is under debate. A modification of the cobalt
average charge would act on our calculations through a modification of
the electrostatic potential seen by the computed cluster. A global
shift of the electrostatic potential seen by the cluster would of
course have no effects on our results. However, a relative shift of
the electrostatic potential seen by the cobalt and oxygen atoms would
have a strong effect. Indeed, a reduction of the potential difference
between the bridging oxygen orbitals and the cobalt magnetic orbitals
can be expected to increase the effective exchange integrals through
the lowering of the ligand-to-metal charge transfer configurations
that mediate the interactions.
On the contrary, an increase of the potential
difference between these orbitals would decrease the effective exchange.

The exclusion effects are treated using total ions
pseudo-potentials~\cite{TIPS} (TIP) on the first shell of atomic sites
surrounding the clusters.

In this description the water molecules are not explicitly
considered. Indeed, we supposed correct the usual idea that their role
is only the increase of the {\bf c} parameter, uncoupling the $\rm
CoO_2$ layers. This point is correctly treated in our calculations
since the atomic positions (both in the quantum clusters and in the
baths) are taken as given in the crystallographic data. 
One could argue that the water molecules should be taken into account
in the clusters environment, however since water do not present a
total net charge but only a dipole moment, its effect is expected to be
very small. We thus neglected this aspect.

The calculations presented in this work are specific for the hydrated
compound. In the dehydrated $\rm Na_{0.35} CoO_2$ phase, the atomic
positions differ from the one used here, and more specifically the
$\rm CoO_2$ inter-layers distances are strongly reduced. Even if the
atomic fractional positions were not changed by the dehydration, the
electrostatic potential seen by the cobalt and oxygen atoms of the
$\rm CoO_2$ layers would be strongly modified and thus the different
orbital energies (and other parameters) would be affected. Whether one
would like to perform calculations for other sodium concentrations,
adequate structural data as well as cobalt average valency should be
taken.

The calculations were done using the MOLCAS~\cite{molcas} and
CASDI~\cite{casdi} set of programs. The basis sets used can be found
in reference~\cite{bases}. The structural parameters were taken from the
Nature paper of Takada~\cite{supra}.

\section{Results}
As mentioned in the previous section we performed two sets of
calculations. The first one aimed at determining the one-site
effective parameters such as the cobalt $3d$ energy splitting due to
the ligand field. This orbital splitting is  renormalized by the
correlation effects within the $3d$ shell as well as by the screening
effects due to the virtual excitations. The second type of
calculations aimed at determining the Co--Co interactions, transfer
between the different $3d$, former $t_{2g}$, orbitals as well as the effective
exchange integrals.

\bigskip

\subsection{The effective on-site $3d$ energy splitting}
\begin{figure*}[ht] 
\resizebox{12cm}{!}{\includegraphics{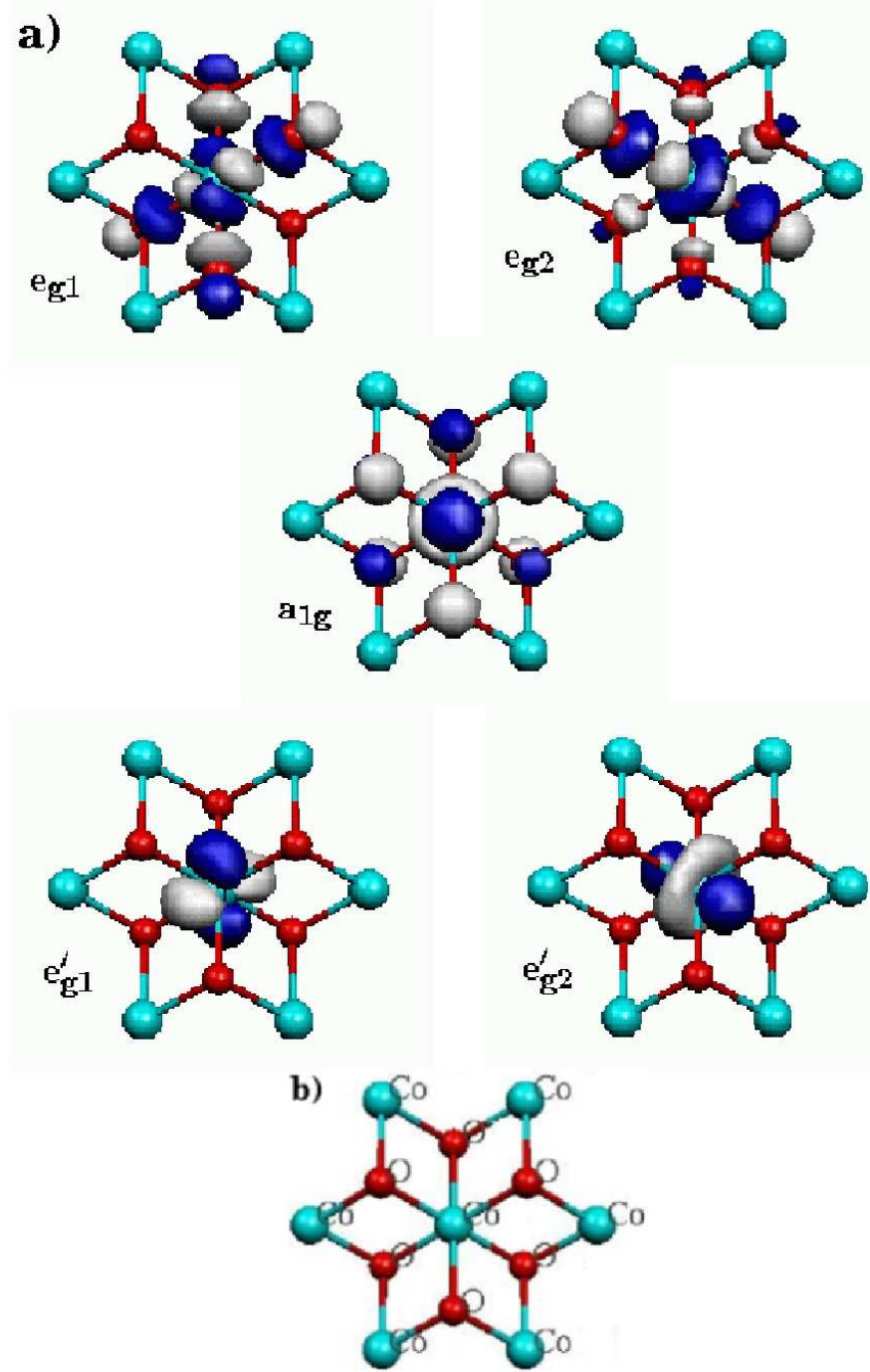}} 
\caption{a) $Co^{4+}$ orbitals  in the $\rm Na_{0.35} Co O_2-1.3H_2O$
compound. b) Label of the different atoms. The ${\bf c}$ axis is orthogonal to the figure plane.} 
\label{f:orb}
\end{figure*}

\begin{figure}[h] 
\resizebox{8cm}{!}{\includegraphics{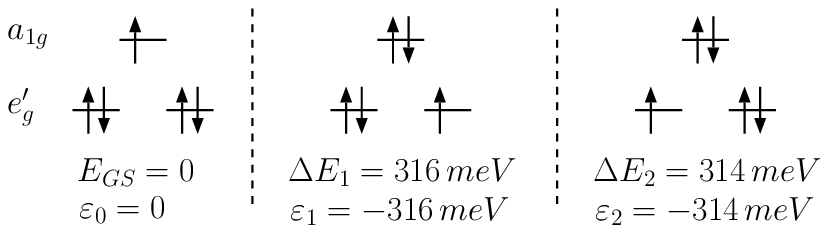}}
\caption{Schematic representation of the ground and first excited states
of the $\rm Co^{4+}$ ions in the $\rm Na_{0.35} Co O_2-1.3H_2O$
compound and corresponding excitation energies.}
\label{f:et2g}
\end{figure}

Calculations with a formal  $\rm Co^{4+}$ cation were performed on the
embedded $\rm Co O_6$ cluster.  Figure~\ref{f:et2g} reports the first
excitation energies and the dominant term of the associated wave
functions. Figure~\ref{f:orb} shows the Co atomic orbitals.  

One sees immediately that the $a_{1g}$ orbital is a $3d_{c^2-r^2}$
orbital in the ${\bf a,b,c}$ crystallographic axes. As stated in many
papers it can be written in as
$$d_{c^2-r^2} = 
\left(d_{xy} + d_{xz}+ d_{yz}\right) / \sqrt{3}$$ if $x,y,z$ are
the Co--O nearly orthogonal directions of the $\rm CoO_6$
octahedron. 
As expected the three low energy orbitals are pointing between the
oxygen atoms, while the high energy ones are directed toward the
ligands.  Another important point to notice is the amount of
Co($3d$)--O($2p$) hybridization in the cobalt orbitals. Indeed, while
the two $e_{g}^\prime$ orbitals do not present noticeable
delocalization on the neighboring oxygen ligands, the $a_{1g}$
orbital exhibits some mixing with the oxygen $2p_z$ orbitals ---~namely about
$2/3$ Co($3d$) and $1/3$ O($2p$)~--- and the two $e_g$ orbitals are
strongly hybridized with the oxygen $2p$ with for the two $e_{g}$ about
$55\%$ Co($3d$) and $42\%$ on the O($2p$).

\begin{figure}
\resizebox{8cm}{!}{\includegraphics{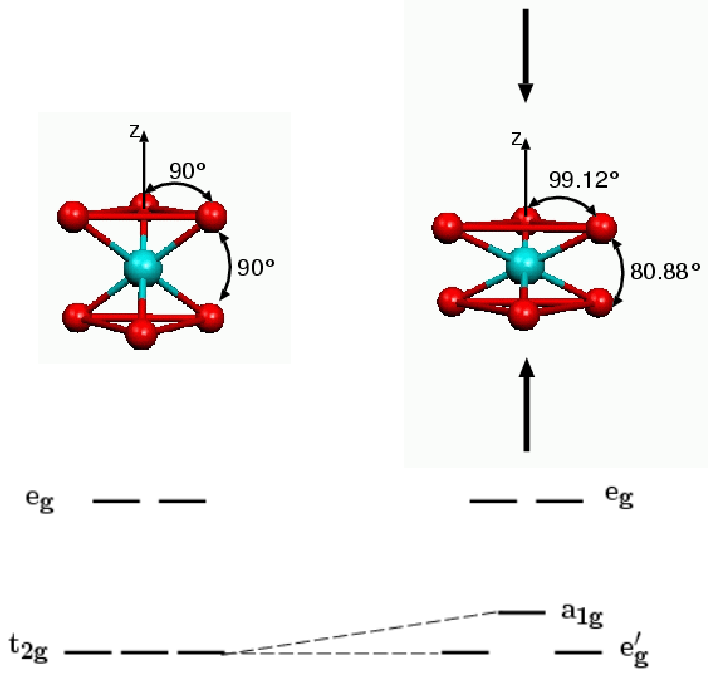}}
\caption{Schematic representation of the trigonal distortion ligand
field splitting as found from the ab initio correlated
calculations. Angles in the distorted structure correspond to the atomic positions given in reference~\cite{supra}.}
\label{f:split}
\end{figure}

Figure~\ref{f:et2g} shows us that in the ground state, the hole is
located on the $a_{1g}$ orbital, which is destabilized compared to the
$e_g^\prime$. Figure~\ref{f:split} pictures this result as the
effective ligand field splitting associated with the trigonal
distortion. This result is in agreement with the finding of the
different DFT ab initio calculations. It however disagrees with the
ligand field analysis of reference~\cite{Mae}. It is well known that
when the transfer (overlap) between a $3d$ metal orbital and the
occupied ligand orbitals increases, the metal $3d$ orbital is
destabilized.  This can be seen in a simple tight binding picture
between a metal $3d$ and ligand $2p$ orbitals. At the second order of
perturbation, the metal $3d$ orbital is destabilized by the quantity
$t^2/(\varepsilon_d - \varepsilon_p)$ while the ligand $2p$ orbital is
stabilized by the same value, due to their hybridization.  This is
presently the case, since the apex oxygens approach the cobalt plane
in the distorted tetrahedron, and thus the overlap between the
$a_{1g}$ cobalt orbital and the oxygen $2p$ orbitals is slightly
augmented.  The computed resulting effective orbital energy splitting
is of the order of $300\, meV$ as can be seen on figure~\ref{f:et2g},
that is in global agreement with ---~even if somewhat larger than~---
the LDA+U estimation~\cite{DFT2} (0.2~eV) from the top of the $a_{1g}$
and $e_g^\prime$ bands.
Finally, one should notice a further splitting between the
$e_g^\prime$ orbitals themselves. This very small splitting is due to
the fact that the electrostatic field generated by the sodium cations
on the cobalt does not present a perfect three-fold symmetry. This
 value is however very small and can be neglected for any practical
purpose.

The $a_{1g}$--$e_g$ splitting can be extracted from the $\rm Co$
atomic $d\longrightarrow d$ excitations of higher energies. Typically
the $S=3/2$ and $S=5/2$ states should be computed. We found a
$a_{1g}$--$e_g$ splitting of $1.8\rm eV$ and thus a
$e_g^\prime$--$e_g$ splitting of $2.1\rm eV$, in global agreement with
the LDA evaluations of $2.5\rm eV$~\cite{picket}.

\subsection{The inter-atomic interactions}
As mentioned in the introduction, one of the open questions in the
modeling of the present compound is whether only the $a_{1g}$ band
is important for the physical properties or whether one should
consider a multi-band model. Indeed, Lee {\it et al}~\cite{picket}
proposed from LDA+U calculations that a crossover occurs between a
single band behavior and a three band behavior as a function of the
band filling. For an equivalent sodium concentration $x<0.5$ they
suggest a three band model and for $x>0.5$ a single band one. On the
model point of view, while several authors argue, on analytical as
well as numerical results, that proper superconductivity behavior
cannot be found using a single band $t-J$ type model~\cite{3bsupra},
other authors found superconducting pairing within a single band $t-J$
model~\cite{1bsupra}. 

In order to address this question one should be able to accurately
evaluate the effective transfer and exchange integrals between the
three former $t_{2g} \longrightarrow 2e_g^\prime + a_{1g}$
orbitals. It is crucial in their evaluation to properly take into
account all Coulomb repulsions, exchange and  screening
effects, as well as the metal--ligands charge transfers.

\begin{figure}[h] 
\resizebox{6cm}{!}{\includegraphics{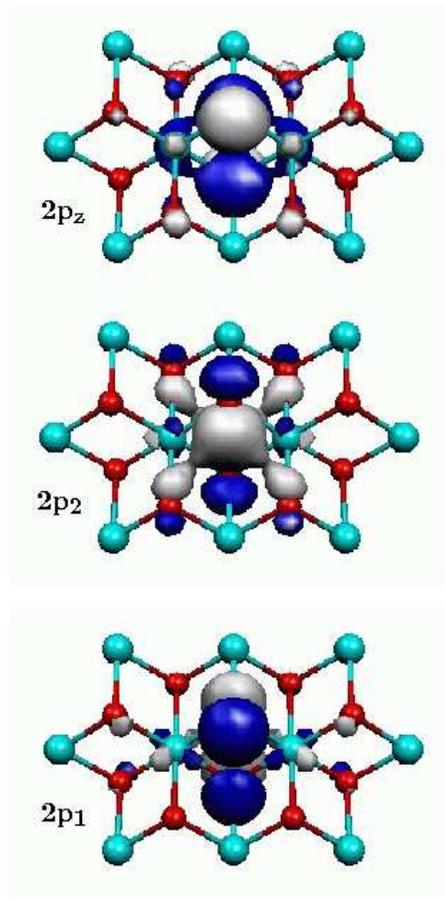}} 
\caption{Oxygen $2p$ orbitals mediating the interactions between the
$a_{1g}$ and $e_g^\prime$ orbitals of the two Co atoms. One sees that
the $p_z$ and $p_2$ orbitals overlap with both the $a_{1g}$ and
$e_{g2}^\prime$ cobalt orbitals, while the $p_1$ overlap with the
$e_{g1}^\prime$ cobalt orbitals.}
\label{f:orbp}
\end{figure}

\begin{figure*}[ht] 
\hspace*{-1cm}\resizebox{14cm}{!}{\includegraphics{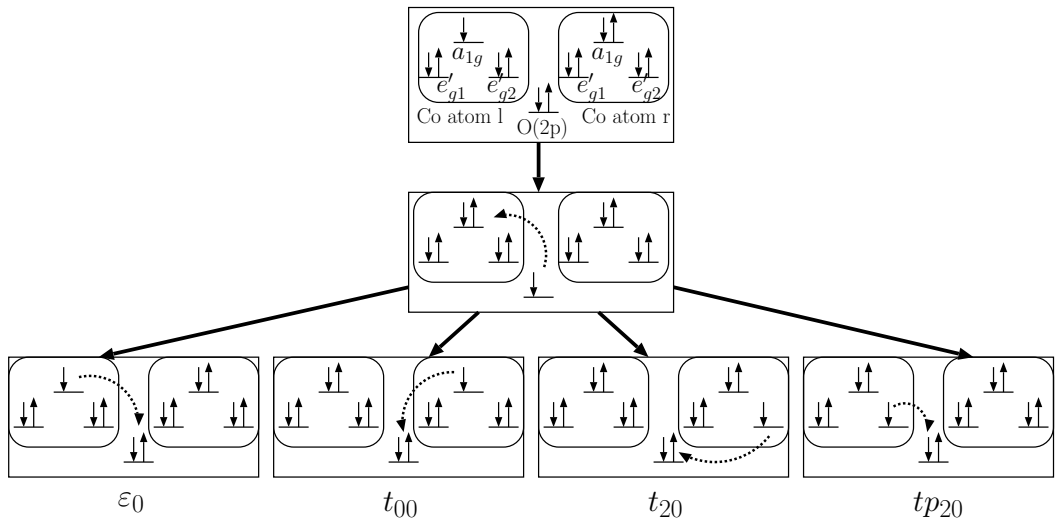}}
\caption{Through bridge perturbative mechanism responsible for
transfers between the $a_{1g}$ and $e_{g2}^\prime$ orbitals of two
nearest neighbor cobalt. The representation of the left and right Co
atoms as well as the orbital denominations are given in the starting
step. The arrows picture at each step the electron transfer yielding
the present configuration. Under each configuration of the last step,
the hopping or energy term to which it contributes is
indicated. Analytic formula associated with the present picture can be
found in the set of equations~\ref{eq:pt} to~\ref{eq:pte}.}
\label{f:tp}
\end{figure*}

These effective transfer integrals between two nearest neighbor cobalt
atoms can be extracted from the first electronic excitations of an
embedded $\rm Co_2 O_{10}$ cluster with $\rm Co^{3+}$--~$Co^{4+}$
mixed cobalt valency.  The computed lowest six states can be associated
with symmetric and antisymmetric combinations of the atomic states
shown in figure~\ref{f:et2g}. As mentioned earlier, the effective
transfer integrals are strongly mediated by the oxygen $2p$
orbitals. Figure~\ref{f:orbp} shows the oxygen $2p$ orbitals bridging the
Co~--~Co interactions.
One can write the following matrix interaction between the
cobalt $a_{1g}$ and $e_g^{\prime}$ $3d$ orbitals and the bridging
oxygen $2p$ ones
$$
 \hspace*{-0.5cm}
\bordermatrix{\vspace*{1ex}
        & a_{1g}^l &e_{g1}^{\prime~l} & e_{g2}^{\prime~l} && 
          a_{1g}^r &e_{g1}^{\prime~r} & e_{g2}^{\prime~r} &&
          2p_z  &2p_1      &2p_2     \cr
& \varepsilon_0 \cr 
& 0 &\varepsilon_1 \cr 
&0 & 0 & \varepsilon_2  
 \vspace*{2ex} \cr
& t^d_{00} &0 & t^d_{20} && \varepsilon_0 \cr 
& 0  & t^d_{11} & 0 &&  0 &\varepsilon_1 \cr 
&  t^d_{20} & 0 & t^d_{22} && 0 & 0 & \varepsilon_2  
  \vspace*{2ex}\cr
& t_{0p_z}& 0 & t_{2p_z} &&  t_{0p_z}& 0 & t_{2p_z} &&  \varepsilon_{p_z} \cr 
& 0 &  t_{1p_1} & 0 &&0 &  t_{1p_1} & 0 && 0  &\varepsilon_{p_1} \cr 
& t_{0p_2} & 0 &  t_{2p_2} &&  t_{0p_2} & 0 &  t_{2p_2} && 0&0& \varepsilon_{p_2} \cr
}       
$$
%
where the $l$ and $r$ superscripts are associated with the two cobalt
atoms, the $\varepsilon_i$ diagonal energies are the effective orbital
energies, $t_{ip_j}$ are the cobalt $3d_i$--oxygen $2p_j$ transfers
and the $t^d_{ij}$ are the direct transfer integrals between the
$3d_i$ and $3d_j$ orbitals of the two cobalt atoms. The direct
integrals are small, however non negligible due to the  short
$\rm Co$--$\rm Co$ distance and the $p$--$d$ hybridization as far as the
$a_{1g}$ orbitals are concerned. In fact the $t_{22}^d$ direct
transfer could be omitted, since this one is really very
small. In addition, we will see later that we cannot explain our
results without explicitly considering at least the $t_{02}^d$
term. Such a $3d-2p$ model may be considered as a bit too complex for
practical uses. In addition, a rapid analysis of the computed wave
functions shows that the explicit contribution of the $\rm{O}(2p)
\rightarrow \rm{Co}(3d)$ excitations are quite small, even if very
important for the mediation of the interactions between the two cobalt
atoms. Indeed, the weight of these configurations is less than $4\%$
in the wave functions.  We can thus reduce the previous matrix into an
effective Hamiltonian on the sole cobalt $3d$ orbitals. All
the effects of the oxygen $2p$ orbitals should however  be taken into account
to properly describe the physics. It results the following effective
Hamiltonian, where both inter-atomic and intra-atomic coupling terms
appear between the $a_{1g}$ and $e_{g2}^{\prime}$ orbitals.
\begin{eqnarray}
 \label{eq:hd}
H_d &=&
\bordermatrix{\vspace*{1ex}
        & a_{1g}^l & e_{g1}^{\prime~l} &e_{g2}^{\prime~l} &&a_{1g}^{r} &
  e_{g1}^{\prime~r}& e_{g2}^{\prime~r}\cr 
        & \varepsilon_0 & 0 & tp_{20} && t_{00} &0 & t_{20} \cr
        & 0 &\varepsilon_1 & 0  && 0  & t_{11} & 0 \cr 
        & tp_{20} & 0 & \varepsilon_2 &&  t_{20} & 0 & t_{22} 
\vspace*{2ex}\cr
        & t_{00} &0 & t_{20} && \varepsilon_0 & 0 & tp_{20} \cr
        & 0  & t_{11} & 0 &&  0 &\varepsilon_1 & 0 \cr
        &  t_{20} & 0 & t_{22} && tp_{20} & 0 & \varepsilon_2 \cr
}       
\end{eqnarray}
where $t_{ij}$ are the effective resulting transfer integrals (direct
 plus mediated by the oxygen ligands) between the $3d_i$ orbital of
 one cobalt and $3d_j$ of the other, $tp_{02}$ is the intra-atomic
 $a_{1g}$--$e_{g2}^\prime$ effective transfer resulting from the
 interactions with the oxygen $2p$ orbitals. This last $tp_{02}$
 integral is in fact quite surprising since one does not expect such
 an intra-atomic effective transfer to take place. It however can
 easily be explained in perturbation theory. Figure~\ref{f:tp}
 pictures the mechanism responsible for the effective transfers
 between the $a_{1g}$ and $e_{g2}^\prime$ orbitals of the two cobalt
 atoms, mediated by one oxygen orbital.
The fact that both the $a_{1g}$ and $e_{g2}^\prime$ orbitals present
non negligible transfer integrals toward the same oxygen $2p$
orbitals result in the fact that four different configurations couple,
at the second order perturbation theory, to the ground state of the
$\rm Co^{4+}$--$\rm Co^{3+}$ ions. Indeed, one has the following
expression for the effective transfers between the $a_{1g}$ and
$e_{g2}^\prime$ orbitals of the Co atoms
\begin{eqnarray} \label{eq:pt}
t_{00} &=& t_{00}^d - {\left(t_{0p_z}\right)^2 \over \Delta_0} 
- {\left(t_{0p_2}\right)^2 \over \Delta_0} \\
t_{20} &=& t_{20}^d - {t_{0p_z}t_{2p_z} \over \Delta_{02}} 
- {t_{0p_2}t_{2p_2} \over \Delta_{02}}\\
tp_{20} &=& - {t_{0p_z}t_{2p_z} \over \Delta_{02}} 
- {t_{0p_2}t_{2p_2} \over \Delta_{02}}
\end{eqnarray}
with $\Delta_0 = \varepsilon_0 - \varepsilon_p + 5(U - 2J_H) - U_p$,
the $\Delta_{02}$ denominator being in the Descloiseaux acception
of the quasi-degenerate perturbation theory~\cite{qpt2} 
$$ {2\over \Delta_{02}} = {1 \over \Delta_0} + {1 \over \Delta_2} $$
with $\Delta_2= \varepsilon_2 - \varepsilon_p + 5(U - 2J_H) - U_p$.\\
$U$ is the Coulomb repulsion for two electrons in the same Co $3d$
orbital, $J_H$ is the Hund exchange, and $U_p$ is the Coulomb
repulsion in the oxygen orbitals.  Finally, the last term corresponds
to a renormalization of the $a_{1g}$ orbital energy
\begin{eqnarray} \label{eq:pte}
\varepsilon_0 &\rightarrow & \varepsilon_0 
- {\left(t_{0p_z}\right)^2 \over \Delta_0} 
- {\left(t_{0p_2}\right)^2 \over \Delta_0} 
\end{eqnarray}
Similarly the perturbation theory yields, at the second order, for
$t_{11}$ and $t_{22}$
\begin{eqnarray*}
t_{11} &=& t_{11}^d - {\left(t_{2p_1}\right)^2 \over \Delta_1} \\
t_{22} &=& t_{22}^d - {\left(t_{2p_z}\right)^2 \over \Delta_2} 
- {\left(t_{2p_2}\right)^2 \over \Delta_2} 
\end{eqnarray*}
with $\Delta_1= \varepsilon_1 - \varepsilon_p + 5(U - 2J_H) - U_p$. \\


If the $tp_{20}$ term is not included in the effective model, it is
impossible to fit the computed wave functions and energies with a good
accuracy. Of course, it is possible to rotate the orbitals in such a
way to nullify the $tp_{20}$ term. It would however result in a set
of orbitals that would be i) bond dependent and ii) no more belonging
to the irreducible representation of the whole system. Indeed, the
axis of the $a_{1g}$ orbital would for instance be tilted with respect
to the {\bf c} axis in a direction dependent of the Co--Co
direction. In order to keep a simple model, respecting the system
symmetry, it appears better to us to use a non-zero $tp_{20}$ term. 


The quality of the model can be evaluated by two ways. The
first criterion is the norm of projection of the computed wave functions
on the configurations space associated with the model (model space). In
the present case the model space is the six combinations of the atomic
electronic states pictured in figure~\ref{f:et2g}. Large norms
warrant that the space supporting the model captures the physics of
the system. In our calculations the minimal value obtained for the
norms of the six projected wave functions is $0.87$. The missing part of
the computed wave function is composed by the excitations responsible
for the screening effects that is more than 17 millions configurations
in our calculation. Table~\ref{t:t} displays the effective transfer
parameters obtained from the fit of $H_d$.
\begin{table}[htbp]   
  \begin{center}
    \begin{tabular}{c|*{5}{c}}
      Parameter & $t_{00}$ & $t_{11}$ & $t_{22}$ & $t_{20}$ & $t_p$ \\
      \hline
      Value (meV) & -276 & 348 & -12 & -89 & -53 
    \end{tabular}
  \end{center}
  \caption{Effective hopping parameters between former $t_{2g} \rightarrow 
  a_{1g} +2 e_g^\prime $ orbitals.}
\label{t:t}
\end{table}
The present fit was done by fulfilling (at the best) the standard
effective Hamiltonians requirements~\cite{heff}~; that is \begin{itemize}
\item that the projection of the exact wave functions in the space
supporting the effective Hamiltonian are eigenfunctions of the later,
\item associated with the exact energies.
\end{itemize}
It results in the minimization over the $H_d$ parameters of the
following Lagrangian
$$ {\cal L}^2 = \sum_m |H_d P \Psi_{exact}(m) -
E_{exact}(m)P\Psi_{exact}(m)|^2 $$ where $P$ is the projection over the
space supporting $H_d$ and $\Psi_{exact}(m)$ are the ab initio wave
functions associated with the ab initio energies $E_{exact}(m)$.  The
quality of the fit is very good since the average error (${\cal L}/6$) can be
evaluated to 1meV.

Let us now concentrate on the effective exchange integral $J$ between
the $a_{1g}$ Fermi level orbitals. $J$ can be evaluated from the
singlet-triplet excitation energy on a $\rm Co_2 O_{10}$ embedded
cluster with two formally $\rm Co^{4+}$ ions. Our calculations yield
an antiferromagnetic coupling, in agreement with the experimental
findings~\cite{afmexp} and LDA+U calculations for large $U$
values~\cite{picket}. The computed value is 
$$J= -66\rm meV$$

One should note that our evaluation of both the effective transfer and
exchange integrals are larger than the LDA and even LDA+U evaluations
found in the literature. This fact is due to the well known problem of
the density functional methods to correctly treat strongly correlated
systems. Indeed, it is well known that in numerous cases, the DFT
results strongly overestimated the ferromagnetism, in particular in
transition metal oxides where $3d$ orbitals play an important
role. This is in particular the case for the present compound since
LDA finds, for all compositions from $x=0.3$ to $x=0.7$, the system
ferromagnetic and metallic, in disagreement with experimental
results. LDA+U somewhat correct this problem, however only in a
mean-field way, the quantum fluctuations due to the electronic
correlation being ignored. In the present case, the authors of
reference~\cite{ldauill} show that the LDA+U method of
incorporating correlation effects is ill-suited for the $Na_xCoO_2$
family of compounds.

\subsection{One site bi-electronic repulsion and Hund's exchange}
The effective Coulomb repulsion of two electrons in the same $a_{1g}$
orbital, $U$, can be extracted from the present calculations in two
different ways.  On one hand, $U$ can be evaluated from the
ground-state energy difference between the $Co^{3+}$ and $Co^{4+}$
ions, embedded in the $\rm Na_{0.35} CoO_2-1.3H_2O$. It yields
$U_1=4.1\,eV$.  On the other hand, $U$ can also be evaluated from the
knowledge of both the hopping and exchange integrals between two
nearest neighbor $a_{1g}$ Co orbitals. Its value can thus be estimated
to $U_2=4.8\,eV$. One immediately notices that the first estimation is
somewhat weaker than the second one. Let us analyze this discrepancy. 
\begin{itemize}
\item The first value ($U_1$) corresponds to a static increase of the
cobalt charge, while the second ($U_2$) corresponds to the energy of
the quantum fluctuations~: $\rm Co^{3+}$--$\rm Co^{5+}$ in a globally
$\rm Co^{4+}$--$\rm Co^{4+}$ state. The Hubbard and related models use
a unique parameter for these two concepts. As far as the raw repulsion
integrals are concerned, there is indeed a unique $U$. However,
screening effects act differently on the static $\rm Co^{3+}$
configuration leading to $U_1$ and on the quantum $\rm Co^{3+}$--$\rm
Co^{5+}$ fluctuations leading to $U_2$. Indeed, it is well known in
quantum chemistry that the static screening acting on $U_1$ is quite
larger that the dynamical one acting on $U_2$. It thus results in a
smaller value for $U_1$ compared to $U_2$.
\item The $U_2$ value corresponds in fact to the difference between
the one site repulsion $U$ and the repulsion $V$ between the $a_{1g}$
orbitals of nearest neighbor $\rm Co$ atoms. It means that the on-site
repulsion responsible for the correct quantum fluctuations between two
nearest neighbor is even larger than the computed $U_2$ value. 
The $4.8eV$ value thus constitutes a lower bound for the repulsion
responsible for the quantum fluctuations.
\end{itemize}
Even-though one cannot definitely conclude on the relative values of
$U$ and $V$, one can expect from the above analysis that the effective
repulsion between nearest neighbor $\rm Co$ sites is probably small.
The above $U_2$ values were extracted for a one-band model. Whether
one would like to use a three band model based on both the $a_{1g}$
and the $e_g^\prime$ orbitals, one should take off the screening
effects due to latters. This can be done in a perturbative
manner. Indeed, the difference in the screening of the $U_2$ parameter
between the one-band and three-bands models can be evaluated to
$$ U_2(1b) = U_2(3b) + \Delta$$
where $U_2(1b)$ is the one-band $U_2$ and $U_2(3b)$ is the three-band
one and 
$$ \Delta = -2{t_p^2 \over \delta -2J_H} 
            +2{t_p^2 \over \delta } 
            -2{t_{20}^2 \over \delta -U}
            +2{t_{20}^2 \over \delta -U-2J_H}
            -2{J_H^2 \over 2\delta }      
$$ 
The on-site repulsion between two different $3d$ orbitals
$U_{dd^\prime}$ was taken in the classical way equal to $U -
2J_H$.  Using the numerical values of the present work we find
$$\Delta = -193 {\rm meV} \qquad {\rm thus} \qquad U_2(3b) = 5.0{\rm
eV} $$

Finally we would like to recall that the
Coulomb repulsion in the $a_{1g}$ orbitals was evaluated from soft
X-ray photo-emission spectroscopy~\cite{uphoto}. The cobalt core $2p$
spectrum exhibits well separated $\rm Co^{4+}$ and $\rm Co^{3+}$
levels, consistent with a repulsion value in the range of $U\sim
3$--$5\, eV$, in total agreement with our calculations.  Let us note
that these values are somewhat weaker than the values usually taken in
LDA+U calculations~\cite{ucal,picket,DFT2} ($5-8\,eV$).

The Hund intra-atomic exchange integral, $J_H$, between the $3d$
orbitals of the cobalt atom can be evaluated from the excitation
energies between the ground state and the higher spin states ($S=3/2$,
$S=5/2$) of the $\rm Co^{4+}$ ion. It comes $J_H= 276\, meV$, to be
compared with the raw $e_{g}^\prime$--$a_{1g}$ exchange integral of
$0.95\, eV$.  Hund's exchange values are thus strongly renormalized by
the screening effects in $\rm Na_{0.35} Co O_2-1.3H_2O$, leading to
the low-spin ground-state observed in the $\rm Co O_2$ layers. Indeed,
the low-spin, high-spin excitation energy can be written as
$2(\varepsilon_{e_g} - \varepsilon_{e_g^\prime}) - 10 J_H$. With a
Hund's exchange of about $1\, eV$, a low spin ground state would
necessitate an $e_g$--$e_g^\prime$ splitting of more than $5\, eV$, in
total disagreement with all experimental and theoretical results.

\section{Discussion and conclusion}
In the present work we determined the effective on-site and coupling
parameters for the $\rm Na_{0.35} Co O_2-1.3H_2O$ compound from ab
initio quantum chemical calculations properly treating both the strong
correlation effects within the cobalt $3d$ shell and the screening
effects on the effective parameters. We determined the ligand field
splitting as well as the on-site Coulomb repulsion and Hund's exchange
within the Co $3d$ orbitals. As far as the interactions between two
cobalt atoms are concerned, we evaluated both the effective transfer
integrals between the $t_{2g}$ orbitals as well as the effective
exchange. It is noticeable that, the ligand field splitting between
the $a_{1g}$ and $e_g^\prime$ orbitals resulting from the splitting of
the $t_{2g}$ orbitals is of the same order of magnitude (and even a
little weaker) as the transfer integrals between two nearest neighbor
Co atoms. In our opinion, both the $a_{1g}$ and the $e_g^\prime$
orbitals should be taken into account in a proper description of this
system.

\acknowledgments The present calculations were done at the CNRS/IDRIS
computational facilities under project n$^\circ$1842. The authors thank
Dr. D. Maynau for providing them with the CASDI suite of programs.


\end{document}